\documentclass[aps,prb,twocolumn,amssymb,floatfix,epsfig]{revtex4}
\usepackage{graphicx,epsfig,psfrag,subfigure,amsopn}
\usepackage{color}
\usepackage{hyperref}
\usepackage{hypcap}
\usepackage{amssymb,amsbsy,amsmath}
\usepackage{amsmath}
\newcommand\beq{\begin{equation}}
\newcommand\eeq{\end{equation}}
\newcommand\bea{\begin{eqnarray}}
\newcommand\eea{\end{eqnarray}}
\newcommand\bsq{\begin{subequations}}
\newcommand\esq{\end{subequations}}

\newcommand\al{\alpha}
\newcommand\be{\beta}
\newcommand\ga{\gamma}
\newcommand\de{\delta}
\newcommand\De{\Delta}
\newcommand\ep{\epsilon}
\newcommand\si{\sigma}
\newcommand\Si{\Sigma}

\newcommand\om{\omega}
\newcommand\ta{\theta}

\newcommand\pa{\partial}
\newcommand\ua{\uparrow}
\newcommand\da{\downarrow}
\newcommand\non{\nonumber}

\newcommand\ig{\includegraphics}
\newcommand\bib{\bibitem}

\begin{document}

\title{Generating surface states in a Weyl semimetal by applying
electromagnetic radiation}

\author{Oindrila Deb and Diptiman Sen}
\affiliation{Centre for High Energy Physics, Indian Institute of Science, 
Bengaluru 560 012, India}

\date{\today}

\begin{abstract}
We show that the application of circularly polarized electromagnetic
radiation on the surface of a Weyl semimetal can generate states at that 
surface. These states can be characterized by their surface momentum. The 
Floquet eigenvalues $e^{i\ta}$ of these states come in complex conjugate pairs 
rather than being equal to $\pm 1$. If the amplitude of the radiation is 
small, we find some unusual bulk-boundary relations: the values of $\ta$ 
of the surface states lie at the extrema of the $\ta$'s of the 
bulk system, and the peaks of the Fourier transforms of the surface state
wave functions lie at the momenta where the bulk $\ta$'s
have extrema. For the case of zero surface momentum, we can analytically
derive scaling relations between the decay length of the surface states 
and the amplitude and penetration length of the radiation. For topological 
insulators, we again find that circularly polarized radiation can generate 
states on the top surface; these states have much larger decay lengths than 
the surface states which are present even in the absence of radiation. 
Finally, we show that radiation can generate surface states for trivial 
insulators also.
\end{abstract}

\maketitle

\section{Introduction}
\label{sec:intro}

Weyl semimetals (WSMs) have emerged as a new class of three-dimensional 
topological materials with gapless and linearly dispersing states in the
bulk~\cite{wan,burkov1,halasz,hosur1,zyuzin1,zyuzin2,son,ojanen,hosur2,
turner,landsteiner,potter,vafek,burkov2,rao,xu1,xu2,lv1,lv2,jia,behrends}.
The low-energy bulk excitations of WSMs are Weyl fermions, 
and the gapless points in the bulk of a WSM are known as Weyl points. 
Each Weyl point is associated with a chirality which is a quantum 
number that depends on the Berry flux enclosed by a surface surrounding
the Weyl point. Hence Weyl points with opposite chiralities act as monopoles 
and anti-monopoles of Berry flux~\cite{turner}. In the presence of 
perturbations, Weyl points can only appear or disappear in pairs which have 
opposite chiralities~\cite{nielsen}. As a result of the non-trivial topology 
in the bulk, WSMs host topologically protected gapless states at the surfaces 
of the system. The momenta of these surface states form certain finite curves 
in the two-dimensional Brillouin zone called Fermi arcs. The ends of the Fermi
arcs on a given surface are given by the projections of the momenta of the 
bulk Weyl points on to that surface. As the momentum of a surface state
approaches one of the end points of a Fermi arc, the decay length of the
surface state into the bulk diverges; hence the surface state merges with a
bulk state at the end point. If the projection of the bulk Weyl
points on a particular surface happens to be a single point, there are 
no states on that surface (see Fig.~\ref{fig:weyl}).

There have been extensive studies in recent years showing that time periodic 
variations of certain parameters in the Hamiltonian of a quantum system can 
drive it into phases which are not present in their static counterparts. 
For example, it has been found that the application of electromagnetic 
radiation (EMR), which is described by a vector potential which varies 
periodically with time, may drive a system from one topological (or 
non-topological) phase to another~\cite{inoue,lindner1,kitagawa1,lindner2,
cayssol,delplace,katan,usaj,wang1,alessio,narayan,yan,saha1,hubener,
zhang,bucciantini} or can significantly alter the properties of a topological 
system~\cite{wang3,chan,kolodrubetz}. 

It has been shown very recently that circularly polarized EMR applied 
to the surface of a WSM can generate novel surface states which can be studied
using the concept of exceptional points in momentum space~\cite{gonzalez}.
However, a detailed understanding of all the properties of such surface states
is not yet available. In this work, we will study the emergence of the 
states on the surfaces of a WSM and a number of properties of these states
using both analytical and numerical methods, without considering the idea 
of exceptional points.

The plan of our paper is as follows. In Sec.~\ref{sec:model1} we will 
consider a simple model of a WSM which has only two Weyl points in the
bulk whose projection on the top surface happens to be a single point;
hence there are no states on that surface in the absence of EMR. We then 
apply circularly polarized EMR on the top surface; this makes the 
Hamiltonian of the system periodic in time, and the system can be studied 
using Floquet theory. We will consider circular polarization (rather than, 
say, linear polarization) so that the Hamiltonian remains invariant under 
rotations in the $x-y$ plane even in the presence of driving. We describe 
such a polarization using a vector potential of the form ${\vec A} (t) = A
(\cos{\om t}, \eta \sin {\om t}, 0)$ where $\om$ is the frequency of the 
radiation, and $\eta =\pm 1$ for right and left circularly 
polarizations respectively. Further, we assume that the radiation amplitude 
decays exponentially as we go away from the top surface into the bulk,
so that $A=A_0 e^{\ga z}$ where $1/\ga$ is the penetration length. [In this
paper we will always take the top surface to be at $z=0$ and the bulk to be 
in the region $z < 0$. Hence $e^{\ga z}$ decays exponentially as we go from 
the top surface into the bulk.] We will study if EMR generates surface states 
that decay along the $- \hat z$-direction (see Fig.~\ref{fig:weyl}). To 
numerically find states that decay along the $\hat z$-direction we introduce 
a one-dimensional lattice along the $\hat z$-direction.
Each lattice point contains two states corresponding to the spin of the 
electron. Due to translational invariance, the momentum of the surface states 
in the plane of the surface, $(k_x,k_y)$, is a good quantum number; this 
reduces the problem to an effectively one-dimensional one in which the surface
momentum simply appears as a parameter. The rotational invariance of our 
problem allow us to choose $k_y=0$ and only $k_x$ to be nonzero in all our 
calculations. In Sec.~\ref{sec:results1} we discuss the results that we 
obtain numerically. We calculate the Floquet operator $U={\cal T} 
e^{-i\int_0^T dt H(t)}$ which evolves the system for one time period 
$T= 2\pi/\om$, and find the eigenvalues and eigenstates of $U$. 
To find the surface states we look at the inverse participation ratio (IPR)
of all the Floquet eigenstates. The states with larger IPR are more 
localized, and we generally find that the states with the largest IPR values
are localized at the top surface. To identify which states are localized at
the surface, we plot the squares of the absolute values of the wave functions 
of these eigenstates and find which states are localized near the top. The 
wave functions give us information about the decay lengths of these surface 
states; these decay lengths depend on the various parameters of the system 
and the radiation. 

In Sec.~\ref{sec:bbc}, we show that there is an unusual kind of bulk-boundary 
correspondence between the surface states and the Floquet eigenstates of the 
bulk system which has periodic boundary conditions in the $\hat z$-direction 
and therefore has the momentum $k_z$ as a good quantum number. To find such a 
correspondence in our system, we numerically find the Floquet eigenvalues 
(FEVs), $e^{i\ta}$, of the bulk states as a function of $k_z$ and compare 
these with the FEVs of the surface states. We find two interesting relations 
between the two sets of $\ta$'s. 
When we plot the $\ta$'s of the states of the bulk system versus their
momentum $k_z$, we discover that the points where the $\ta$'s have extrema
coincide with the values of $\ta$ of the surface states. We also find
that the peaks of the Fourier transform of the wave functions of the surface 
states occur at the same momenta $k_z$ where the bulk FEVs have extrema.

In Sec.~\ref{sec:zero} we discuss a special case where the momenta of the
surface states, $k_x$ and $k_y$, are both equal to zero. We show that the 
equations of motion of the Floquet problem can be made completely 
time-independent by a unitary transformation. Thus the problem of finding 
the FEVs can be reduced to finding the eigenvalues of a time-independent 
Hamiltonian; these can be easily found numerically. In the continuum limit, 
the equations of motion near the top surface can be written as a differential 
equation of the Airy form which is the Schr\"odinger equation of a particle 
with a hard wall at the top surface and a linearly increasing potential in 
the bulk. We therefore obtain a number of surface state whose wave functions 
look like Airy functions. Further, we can analytically understand some 
features of the surface states such as the dependence of their decay lengths 
on the radiation amplitude.

In Sec.~\ref{sec:topo} we discuss the case of a three-dimensional
topological insulator (TI) and the effect of 
applying circularly polarized EMR near the top surface of such a system. 
The analysis proceeds along similar lines as in the WSM system. We 
numerically show that new surface states are generated by the EMR; these
are in addition to the old surface states which exist even in the absence 
of radiation simply because the system is a TI. The decay lengths of the new
surface states are much larger than that of the old states, and can
be tuned by varying the amplitude of the EMR. Finally, we find that EMR 
can generate surface states even in a trivial insulator which has no surface 
states in the absence of the radiation. In Sec.~\ref{sec:con}, we summarize 
our main results and point out some directions for future studies.

\section{Model}
\label{sec:model1}

We begin with a simple model for the low-energy Hamiltonian of a 
three-dimensional WSM. Expanding around the center of the Brillouin zone to 
quadratic order in the momentum ${\vec k}$, we consider the form
\bea H &=& [m_0 ~-~ m_1 k_z^2 ~-~ m_2 ~(k_x^2+k_y^2)] ~\si^z \non \\
&& +~ v ~[\si^x k_x ~+~ \si^y k_y] \label{ham1} \eea
where $\si^i$, $i=x,y,z$ are the Pauli spin matrices. (We will set $\hbar =1$ 
in this paper). The dispersion for this Hamiltonian is given by 
\bea E_b &=& \pm \sqrt{[m_0-m_1 k_z^2 -m_2 (k_x^2+k_y^2)]^2 + v^2 (k_x^2 + 
k_y^2)} \label{dis} \non \\
\eea
The gapless points of this dispersion define the Weyl points. From 
Eq.~\eqref{dis} we see that $E_b=0$ implies
\bea && k_x ~=~ k_y ~=~ 0, \non \\
&& k_z ~=~ \pm ~\sqrt{\frac{m_0}{m_1}}. \eea
Hence this system has only two Weyl points which lie at $(k_x,k_y,k_z) = 
(0,0,-\sqrt{m_0/m_1})$ and $(0,0,\sqrt{m_0/m_1})$. 
 
\begin{figure}
\epsfig{figure=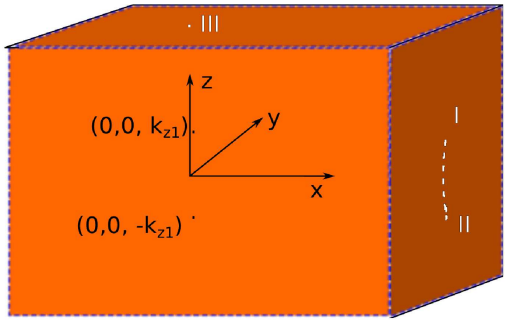,width=7cm}
\caption{The figure shows projections of the two Weyl points on the $x-y$ 
(top) and $y-z$ (side) surfaces.} \label{fig:weyl} \end{figure}

We now look at the projections of the Weyl points on different surfaces of the
three-dimensional system. The projections of the Weyl points on a side surface 
(say, the $y-z$ surface) define a Fermi arc on that surface which, in the 
Brillouin zone of that surface, is a curve going from $(k_y,k_z) = 
(0,-\sqrt{m_0/m_1})$ to $(0,\sqrt{m_0/m_1})$. Hence surface 
states exist on that surface with momenta which lie on the Fermi arc. 
However the projections of the Weyl points on the top surface (the $x-y$ 
surface) go to a single point given by $(k_x,k_y) = (0,0)$, as illustrated in 
Fig.~\ref{fig:weyl}. Hence there is no Fermi arc and therefore no surface 
states at the top surface. 

We will now study what happens if the top surface is illuminated by circularly 
polarized EMR with frequency $\om$ and time period $T=2\pi/\om$.
In the case of circular polarization incident on the $x-y$ surface, the 
radiation produces the vector potential ${\vec A} (t) = A( \cos{\om t},\eta 
\sin {\om t}, 0)$ where $\eta =\pm 1$ for right and left circularly polarized 
beams, respectively. [Actually, since the radiation describes a wave 
traveling towards the $-{\hat z}$ direction, the arguments of $\cos$ and 
$\sin$ in ${\vec A} (t)$ should be $\om t + k z$, where 
$k = \om/c$. However, for the system parameters that we will consider below, 
$\om z/c$ will be much smaller than 1. We will therefore ignore the term $kz 
= \om z/c$.] We want to see if this periodic driving produces any states on 
the $x-y$ surface. To study this problem numerically, we introduce a 
one-dimensional (1D) lattice along the $\hat z$-direction and a continuum 
model along the $\hat x$ and $\hat y$ directions; hence $k_x$ and $k_y$ are 
good quantum numbers. We will use Floquet theory in order to compute the FEVs
as a function of $(k_x,k_y)$, and we will then find the states which are 
localized near the $x-y$ surface using the inverse participation ratio
as discussed below. 

We introduce a lattice discretization of of the $z$ part of Eq.~\eqref{ham1} 
as follows. We label the sites of the 1D lattice by integers $n$, with the 
total number of lattice points being equal to $N_z$; we will use open
boundary conditions, with $n =1$ and $N_z$ denoting the top and bottom
surfaces respectively. Each site will have two states corresponding to spin,
$\si^z = \pm 1$; hence the wave function at each site $n$ will have two 
components, and the entire wave function will have $2N_z$ components. 
We will denote the lattice spacing by $a$ which we will take to be equal to
1 \AA. 
We now define a lattice Hamiltonian which reduces to Eq.~\eqref{ham1} in the 
limit $k_z a \to 0$. This can be done by replacing $k_z^2 \to (2/a^2 )[1 - 
\cos(k_z a)]=(1/a^2)(2-e^{ik_za}-e^{-ik_za})$, and hence $k_z^2 \psi_n 
\to (1/a^2) (2 \psi_n - \psi_{n+1} - \psi_{n-1})$ on the lattice. 
Equation~\eqref{ham1} then takes the form
\bea H\psi_n &=& [m_0 ~-~ m_2 (k_x^2+k_y^2)] ~\si^z ~\psi_n \non \\
&& + ~\frac{m_1}{a^2} ~\si^z ~[\psi_{n+1} ~+~ \psi_{n-1} ~-~ 2\psi_n] 
\non \\
&& + ~v ~[\si^x k_x ~+~ \si^y k_y] ~\psi_n. \label{ham2} \eea
We will set $a=1$ \AA~ henceforth.

Next, we include the vector potential of the circularly polarized EMR in the 
Hamiltonian. We assume that the amplitude of the vector potential has the form 
$A=A_0 e^{-\ga a n}$, so that the amplitude of the vector potential 
decreases exponentially with a penetration length $1/\ga$ as we move away 
from the $x-y$ surface and go into the bulk. It is realistic to assume this 
kind of exponential dependence as radiation usually gets absorbed in a medium.
Adding the vector potentials to the momenta as $k_x \to k_x + A \cos \om t$ and
$k_y \to k_y + A \sin \om t$, we see that Eq.~\eqref{ham2} takes the form
\bea H\psi_n &=& [m_0 ~-~ m_2 \{ (k_x+A\cos{\om t})^2 \non \\
&& ~~~~~~~~~~~~~~~+(k_y+A \sin{\om t})^2 \} ] \si^z \psi_n \non \\
&& + v ~[\si^x (k_x+A\cos{\om t}) + \si^y (k_y+A \sin{\om t})] \psi_n 
\non \\ 
&& + ~m_1 \si^z [\psi_{n+1}+\psi_{n-1}-2 \psi_n], \label{ham3} \eea
where we have assumed that the radiation is right circularly polarized ($\eta 
= 1$).

We will numerically study the eigenvalues and eigenvectors of the Floquet
operator $U$ which time evolves the system by one time period $T= 2 \pi/
\om$, namely, 
\beq U ~=~ {\cal T} e^{-i \int_0^T ~dt H(t)}, \label{flo1} \eeq
where $\cal T$ denotes the time-ordered product.
For a periodic Hamiltonian as in Eq.~\eqref{ham3} which satisfies 
$H (t +T) = H(t)$, the Floquet operator has the property that 
its eigenvalues remain invariant if we shift the time by an arbitrary amount, 
i.e., $t \to t + \tau$. To see this, let us take $\tau$ to lie in the range
$[0,T]$, and consider the Floquet operator corresponding to the time-shifted
problem,
\bea U' &=& {\cal T} e^{-i\int_\tau^{T+\tau} ~dt H(t)} \non \\ 
&=& {\cal T} e^{-i\int_T^{T+\tau} ~dt H(t)} e^{-i\int_\tau^T ~dt H(t)} \non \\
&=& {\cal T} e^{-i\int_0^\tau ~dt H(t)} e^{-i\int_\tau^T~ dt H(t)}, 
\label{flo2} \eea
where we have used the fact that $H(t)$ is periodic in time with period $T$.
If we define $U_1=e^{-i\int_0^\tau dt H(t)}$ and $U_2= e^{-i\int_\tau^T dt 
H(t)}$, we see that $U=U_2 U_1$ while $U'=U_1 U_2$.
Let $\chi$ be an eigenstate of $U$, with $U \chi = e^{i \ta} \chi$ (the
eigenvalues of $U$ are unimodular since $U$ is unitary). This implies that
\bea U_2 U_1 \chi &=& e^{i \ta} ~\chi, \non \\
U_1 U_2 U_1 \chi &=& e^{i \ta} ~U_1 \chi, \non \\
U' (U_1 \chi) &=& e^{i \ta} ~(U_1 \chi). \label{flo3} \eea
Hence, if $\chi$ is an eigenstate of $U$ with eigenvalue 
$e^{i \ta}$, then $U_1 \chi$ is an eigenstate of $U'$ with the same 
eigenvalue. Now, $U_1=e^{-i\int_0^\tau dt H(t)}$ gives the
time evolution from $t=0$ to $t=\tau$. Thus if $\chi$ is the wave function at 
$t=0$, then $U_1 \chi$ is the time evolved wave function at time $t=\tau$. 

We will now use the above facts to show that although Eq.~\eqref{ham3}
contains two momenta $k_x$ and $k_y$, the eigenvalues of the Floquet
operator $U$ only depends on the magnitude $k = \sqrt{k_x^2 + k_y^2}$.
To show this, let us assume that $k_x = k \cos \phi$ and $k_y = k \sin \phi$, 
and carry out a rotation by angle $\phi$ around the $z$-axis in the spin space,
\bea \si'^x &=& \si^x \cos \phi + \si^y \sin \phi, \non \\
\si'^y &=& \si^y \cos \phi - \si^x \sin \phi. \eea
We now look at the various terms in Eq.~\eqref{ham3}. The terms which are 
linear in $(k_x, k_y)$ and $(\si^x, \si^y)$ are given by
\bea \si^x k_x + \si^y k_y &=& ~(\si'^x \cos \phi - \si'^y \sin \phi ) ~k 
\cos \phi \non \\
&& +~(\si'^x \sin \phi + \si'^y \cos \phi ) ~k \sin \phi \non \\
&=& \si'^x k, \non \\
k_x \cos \om t + k_y \sin \om t &=& k \cos (\om t - \phi), \non \\
\si^x \cos \om t + \si^y \sin \om t &=& \si'^x \cos (\om t - \phi) + \si'^y 
\sin (\om t - \phi). \non \\
&& \label{rot} \eea
Hence Eq.~\eqref{ham3} takes the form
\bea H\psi_n &=& [m_0 ~-~ m_2 \{ k^2 + A^2 \non \\
&& ~~~~~~+ 2 k A \cos (\om t - \phi)\}] \si^z \psi_n \non \\
&& + v ~[\si'^x \{k +A\cos (\om t - \phi) \} \non \\
&& ~~~~~+ \si'^y A \sin (\om t - \phi)] \psi_n \non \\
&& + ~m_1 \si^z [\psi_{n+1}+\psi_{n-1} -2 \psi_n]. \label{ham4} \eea
We can now use the arguments presented around Eqs.~(\ref{flo2}-\ref{flo3}) to
show that the eigenvalues of the Floquet operator of Eq.~\eqref{ham4} 
do not depend on the value of $\phi$. It is therefore sufficient to
set $\phi = 0$ in Eq.~\eqref{ham4} and study the FEVs as a 
function of $k$ alone. Namely, we can take $k_x = k$ and $k_y = 0$
in Eq.~\eqref{ham3}.

Using the fact that the Pauli matrices $\si^x$ and $\si^z$ are real while 
$\si^y$ is imaginary, we can derive a symmetry of the Hamiltonian in 
Eq.~\eqref{ham3} and therefore of $U$. We define a $2N_z$-dimensional matrix 
\beq \Si^y ~=~ I_{N_z} ~\otimes~ \si^y, \eeq
where $I_{N_z}$ is the $N_z$-dimensional identity matrix. Namely, $\si^y$
has $N_z$ blocks along the diagonal, each block being equal to $\si^y$.
We note that $(\Si^y)^2 = I_{2N_z}$. We then see that the Hamiltonian in 
Eq.~\eqref{ham3} satisfies
\beq \Si^y ~H~ \Si^y ~=~ - H^*, \label{sym0} \eeq
which implies that the Floquet operator in Eq.~\eqref{flo1} satisfies
\beq \Si^y ~U~ \Si^y ~=~ U^*. \label{sym1} \eeq
This implies that if $U \psi_1 = e^{i\ta} \psi_1$, then 
\beq \psi_2 ~=~ \Si^y \psi_1^* \label{sym2} \eeq 
satisfies $U \psi_2 = e^{-i\ta} \psi_2$.

\section{Numerical results}

\subsection{Surface states}
\label{sec:results1}

In this section we present numerical results for the WSM under 
the effect of the periodic driving. We calculate the Floquet operator $U$ 
defined in Eq.~\eqref{flo1} and find its eigenvectors $\psi_j$ and eigenvalues
$e^{i \ta_j}$, where the $\ta_j$ lie in the range $[-\pi,\pi]$. [The Floquet 
eigenvalues are sometimes written as $e^{i\ta_j} = e^{-i \ep_j T}$, where 
$\ep_j$ are called the quasienergies; these lie in the range $[-\pi/T,\pi/T]$.
However, we will generally work with the variable $\ta_j$ rather than $\ep_j$ 
due to the simplifying feature that the range of $\ta_j$ does not 
depend on $T$.] To identify the surface states, we calculate
the inverse participation ratios (IPRs) of all the eigenvectors. The IPR for a 
state $\psi_j$ is defined as $I_j = \sum_{m=1}^{2N_z} |\psi_j (m)|^4$ (note 
that the states are normalized so that $\sum_{m=1}^{2N_z} |\psi_j (m)|^2 = 1$
for all $j$). States with large values of $I_j$ are more localized; we
look at the wave functions of such states to find which of them are
localized near the top surface ($n = 1$).

\begin{figure}[h]
\epsfig{figure=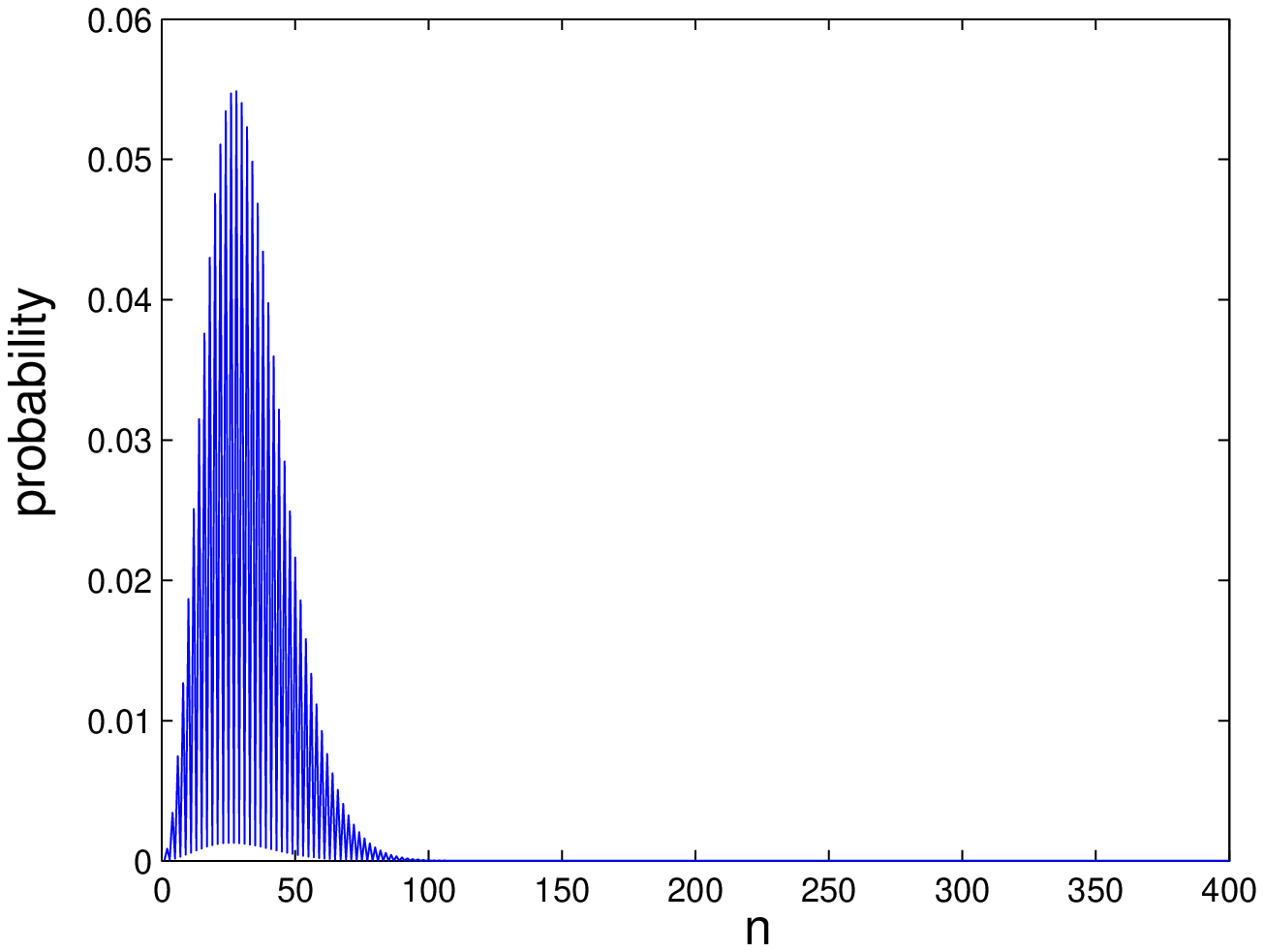,width=8cm}
\caption{Probabilities vs $n$ for a surface state wave function for a 
system with $m_0=0.5$ eV, $m_1= 0.605$ eV, $m_2=1$ eV \AA$^2$, $v=1$ eV \AA, 
$k=0.5$ \AA$^{-1}$, $A_0=0.5$ \AA$^{-1}$, $\ga=0.0005$ \AA$^{-1}$, $\om=2$ 
eV$/\hbar$, and 200 sites. The FEV of this state is $0.0363+ 0.9993 i$.} 
\label{fig02} \end{figure}

\begin{figure}[h]
\epsfig{figure=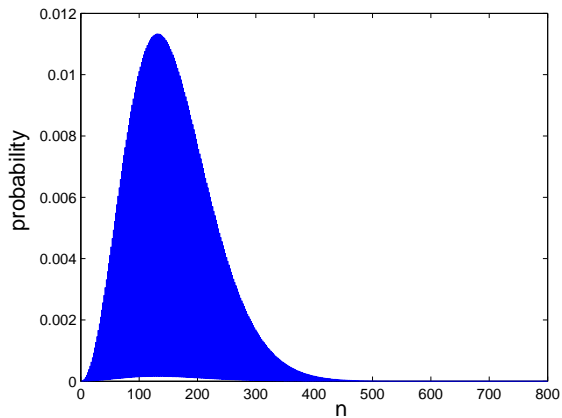,width=8cm}
\caption{Probabilities vs $n$ for a surface state wave function for 
$A_0=0.05$ \AA$^{-1}$, and 400 sites; all other parameters have the same 
values as in Fig.~\ref{fig02}. The FEV of this state is $0.7521+ 0.6590 i$.} 
\label{fig03} \end{figure}

The values of the parameters that we have used to find surface states are as 
follows~\cite{gonzalez}: $m_0=0.5$ eV, $m_1=0.605$ eV, $m_2=1$ eV \AA$^2$, 
$v=1$ eV \AA, $k_x=k=0.5$ \AA$^{-1}$, and $k_y=0$. In Fig.~\ref{fig02}, we show
the probabilities $|\psi_j (m)|^2$ versus $m$ for one of the surface states, 
taking $A_0=0.5$ \AA$^{-1}$, $\ga=0.0005$ \AA$^{-1}$, $\om=2$ eV$/\hbar$, and 
$N_z = 200$ sites (hence the Hamiltonian is a $400$-dimensional matrix). 
Since $A=A_0 e^{-\ga a n}$ (where $a = 1$ \AA~ is the lattice spacing), 
the penetration length of the radiation is $1/\ga = 2000$ \AA~ which is much 
larger than the system size of $N_z a = 200$ \AA; this means the radiation 
decreases very little as we move away 
from the $x-y$ surface into the bulk. In Fig.~\ref{fig03}, we show the 
probabilities of a surface state for $A_0=0.05$ and $N_z = 400$ sites;
all other parameters are the same as in Fig.~\ref{fig02}. We find that
the surface state has a bigger width. We will see later that the width of 
the surface state scales as an inverse power of the radiation amplitude $A_0$.
(We find that it is necessary to give $\ga$ a small but nonzero value to 
obtain surface states. If we set $\ga = 0$ exactly, we do not find any 
states localized near the surface).

Looking at the wave functions of the surface states, we find that 
two surface states $\psi_1$ and $\psi_2$ whose FEVs are 
complex conjugates of each other are related to each other as
\beq \psi_2 ~=~ \si^y \psi_1^* \label{sym3}, \eeq 
as expected from Eq.~\eqref{sym2}.

We find numerically that states appear at the top surface for a range of 
momenta $k_x = k$ and $k_y = 0$, where $k$ goes from, say, $k_1$ to $k_2$. 
This range depends on the various parameters of the model and the EMR.
For example, for the parameters $m_0, ~m_1, ~m_2, ~v, ~A_0, ~\ga$ and 
$\om$ used in Fig.~\ref{fig02}, we find 
that there are at least two surface states (with complex conjugate FEVs)
all the way from $k = 0$ to $0.7$; hence $k_1=0$ and $k_2 = 0.7$. (For certain
values of $k$, there are four or more surface states). Since the problem is 
rotationally invariant, this implies that there is an annular region of 
momentum $(k_x,k_y)$ given by $k_1 < \sqrt{k_x^2 + k_y^2} < k_2$ where 
there are two or more surface states. This region has a finite area given by
$\pi (k_2^2 - k_1^2)$. Now consider a system whose top surface has a very 
large but finite area given by $\int dx dy = \cal A$. Then the total number 
of points in
phase space, defined as $\int dk_x dk_y dx dy/(2\pi)^2$, where surface states
appear is given by ${\cal A} (k_2^2 - k_1^2)/(4 \pi)$. Hence the number of 
surface states is proportional to the surface area ${\cal A}$. Note however 
that the FEVs of these states depend on the value of $\sqrt{k_x^2 + k_y^2}$;
hence the degeneracy of the FEVs is not proportional to the 
area~\cite{gonzalez}.

It is interesting to study how the surface states evolve with time.
Namely, given a surface state $\psi (0)$ which is an eigenstate of the 
Floquet operator in Eq.~\eqref{flo1}, we can study how
\beq \psi(\tau) = {\cal T} e^{-i \int_0^\tau ~dt H(t)} ~\psi (0) 
\label{psit} \eeq
changes with $\tau$ in the range $0 \le \tau \le T$. [The arguments presented
around Eqs.~(\ref{flo2}-\ref{flo3}) show that $\psi (\tau)$ is the eigenstate 
of the time-shifted Floquet operator $U' = {\cal T} e^{-i \int_\tau^{T+\tau}
dt H(t)}$ discussed in Eq.~\eqref{flo2}.] We have 
numerically studied how the wave function in Eq.~\eqref{psit} varies with 
$\tau$. We find that the variation is generally quite small. For instance, 
for the system parameters used in Fig.~\ref{fig03}, the plots of probabilities 
versus $n$ for $\psi(0)$, $\psi(T/4)$, $\psi(T/2)$ and $\psi(3T/4)$ are 
found to be indistinguishable from each other on the scale of that figure.

We have examined if the surface states discussed in this section are stable 
to certain perturbations. We find that these states continue to exist 
when the perturbations are sufficiently small. For instance, we have studied 
the effect of a small random on-site potential (proportional to the 
$2 \times 2$ identity matrix) added to the Hamiltonian in Eq.~\eqref{ham3};
the on-site potential $\ep (n)$ is taken from a uniform probability 
distribution in the range $[-\ep_0,\ep_0]$, and is chosen independently
at different sites $n$. (Such a term in the Hamiltonian breaks the
symmetry described in Eq.~\eqref{sym0}, and hence the eigenvalues of the
Floquet operator $U$ no longer come in complex conjugate pairs). 
Numerically we find that the surface states survive this perturbation, 
possibly with some changes in the shapes and peak positions of their 
wave functions, when $\ep_0 \lesssim 0.1 ~v A_0$.

\subsection{Bulk-boundary correspondence}
\label{sec:bbc}

We now consider the system with periodic boundary conditions in the 
$\hat z$-direction (namely, the site after $n = N_z$ is the same as the 
site at $n = 1$). This gives us a bulk system which has no boundaries. We 
then look at the bulk states in the presence of radiation and see if there is 
any correspondence between the bulk states and the surface states that appear 
due to driving. In the previous section on surface states, we had taken the 
amplitude $A$ of the vector potential to decrease exponentially along the 
$\hat z$-direction as $A=A_0 e^{- \ga a n}$. We took $\ga$ to be very small, 
and the system size $N_z a$ to be much smaller than the penetration length 
$1/\ga$ but much larger than the width of the surface states. As we will see 
later, the width of the surface states scales in such a way with $1/\ga$ that
the surface states survive in the limit $\ga \to 0$ provided that we
take the system size $N_z$ to be large enough. We will therefore
set $\ga = 0$ in this section so that the radiation amplitude is
uniform through out the system. Hence we have translational invariance 
along the $\hat z$-direction and the momentum $k_z$ is a good quantum number,
along with $k_x$ and $k_y$. Setting $k_x = k$ and $k_y = 0$ as
before, the bulk Hamiltonian is found to be
\bea H_b &=& [m_0 ~-~ 2 m_1 (1 - \cos k_z) \non \\
&& ~- ~m_2 ~(k^2 ~+~ A_0^2 ~+~ 2 k A_0 \cos \om t )]~ \si^z 
\non \\
&& +~v ~[\si^x (k + A_0 \cos{\om t}) ~+~ \si^y A_0 \sin \om t]. \non \\
&& \label{ham5} \eea
This Hamiltonian is only a $2$-dimensional matrix unlike the Hamiltonian
in the previous section which was $2N_z$-dimensional; hence the numerical
calculations are much faster here. We find the eigenvalues of the Floquet 
operator $U={\cal T} e^{-i \int_0^T dt H_b(t)}$ as a function of $k_z$, 
for the values of the various parameters for which we got surface states 
in the previous section. [In analogy with the symmetry discussed in 
Eqs.~(\ref{sym0}-\ref{sym2}), we find here that $\si^y H \si^y = - H^*$
which implies that $\si^y U \si^y = U^*$; hence the eigenvalues of $U$
are complex conjugates of each other for every value of $k_z$. It is
also clear from Eq.~\eqref{ham5} that the eigenvalues of $U$ are the
same for $k_z$ and $-k_z$.]

\begin{figure}[h]
\begin{center}
\subfigure[]{\ig[width=8cm]{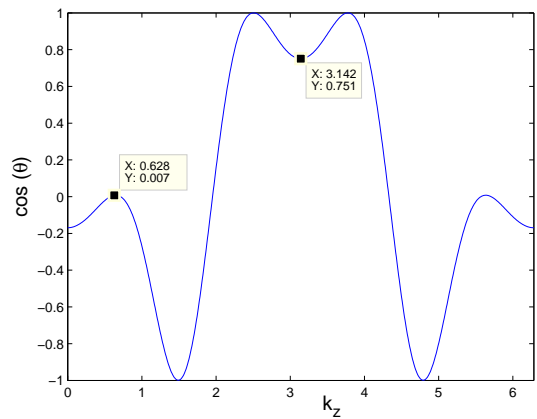}} \\
\subfigure[]{\ig[width=8cm]{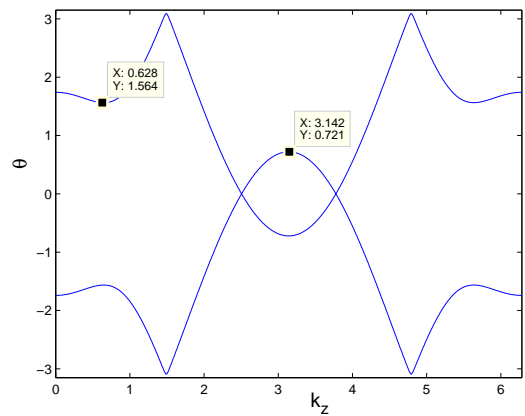}}
\end{center}
\caption{(a) Plot of $\cos (\ta)$ vs $k_z$ and (b) plot of $\ta$ vs
$k_z$ for a bulk system with $m_0=0.5$ eV, $m_1= 0.605$ eV, $m_2=1$ eV 
\AA$^2$, $v=1$ eV \AA$^{-1}$, $k=0.5$ \AA$^{-1}$, $A_0=0.05$ \AA$^{-1}$, 
$\om=2$ eV$/\hbar$, and 800 sites.} \label{fig:bulk} \end{figure}

In Fig.~\ref{fig:bulk} (a), we plot the real part, $\cos \ta$, of the FEVs 
($e^{i\ta}$) as a function of $k_z$ for a bulk system with $m_0=0.5$ eV, 
$m_1= 0.605$ eV, $m_2=1$ eV \AA$^2$, $v=1$ eV \AA$^{-1}$, $k=0.5$ \AA$^{-1}$, 
$A_0=0.05$ \AA$^{-1}$, $\ga = 0$, $\om=2$ eV$/\hbar$, and $N_z = 800$ sites
with periodic boundary conditions. 
We see that $\cos \ta$ has extrema at several values of $k_z$. Of these, the 
extrema at $\cos \ta = \pm 1$ are trivial in the sense that they arise simply 
because $\cos \ta$ lies in the range $[-1,1]$. The non-trivial extrema of 
$\cos \ta$ lie at $k_z = 0.628$ and $2\pi -0.628$ where $\cos \ta = 0.007$ 
and at $k_z = \pi$ where $\cos \ta = 0.754$. As an alternative way of viewing 
these results, Fig.~\ref{fig:bulk} (b) shows a plot of $\ta$ versus $k_z$ for 
the same system parameters. (For each value of $k_z$, there are two values of 
$\ta$ given by a $\pm$ pair). Once again we see some extrema lying at
$k_z = 0.628$ and $2\pi -0.628$ where $\ta = 1.564$ and at $k_z = \pi$ 
where $\ta = 0.721$. 

We now compare the extrema of $\ta$ for the bulk system with the values of 
$\ta$ for 
the surface states studied in the previous section for the 1D lattice system 
with the same parameter values (except that the previous section requires a 
small nonzero value of $\ga=0.0005$ \AA$^{-1}$). We find that the $\ta$ of 
the surface states match some of the extrema of $\ta$ of the bulk system. 
(For instance, the surface state shown in Fig.~\ref{fig03} has $\cos \ta =
0.7521$ implying $\ta=0.720$; this is very close to the extremum value of 
$0.721$ that we see in Fig.~\ref{fig:bulk} (b).) This is one kind of 
bulk-boundary correspondence.



\begin{figure}[h]
\epsfig{figure=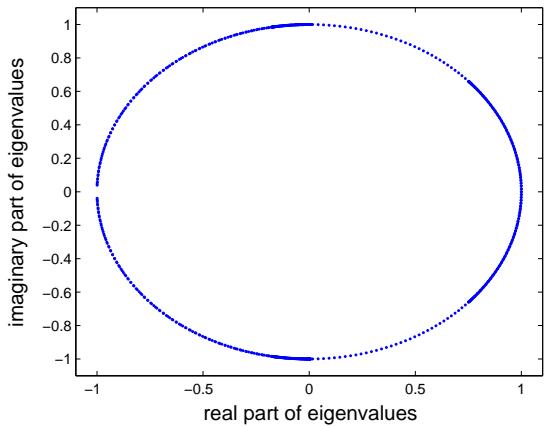,width=8cm}
\caption{Plot of real and imaginary parts of the FEVs of the states of a bulk 
system with the same parameter values as in Fig.~\ref{fig:bulk}.} 
\label{fig:evalue2} \end{figure}

\begin{figure}[h]
\epsfig{figure=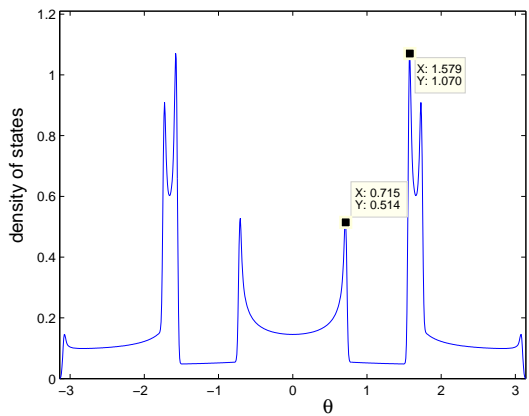,width=8cm}
\caption{Density of states $\rho (\ta)$ as a function of $\ta$ in the range 
$[-\pi,\pi]$ for a bulk system with the same parameter values as in 
Fig.~\ref{fig:bulk}.} \label{fig:dos} \end{figure}

The fact that the FEVs of the bulk system have extrema as a function of $k_z$ 
can be visualized in a different way. Figure~\ref{fig:evalue2} shows the real 
and imaginary parts of the FEVs (these form a unit circle since the Floquet 
operator $U$ is unitary) for a system with the same parameters as in
Fig.~\ref{fig:bulk}. We can see that the density of the eigenvalues is 
not uniform throughout the plot; it is denser in some parts compared to the 
other parts. Near any point $\ta'$ where $\ta_{k_z}$ has an extremum, the 
density of states defined as
\beq \rho (\ta) ~=~ \int_0^{2\pi} ~\frac{dk_z}{2\pi} ~\de (\ta - \ta_{k_z}) 
\label{dos} \eeq 
will diverge for an infinitely large system and will have a sharp peak for a 
finite but large system. (In Eq.~\eqref{dos} we have normalized the density of 
states such that $\int_{-\pi}^\pi d \ta \rho (\ta) = 1$). In 
Fig.~\ref{fig:dos} we show the density of states as a function of $\ta$.
[In order to obtain a smooth curve, we have replaced the $\de$-functions in 
Eq.~\eqref{dos} by Gaussians $\exp[-(\ta - \ta_{k_z})^2/\si^2]$, where $\si 
= 0.0224$, and then summed over the $2N_z$ values of $\ta_{k_z}$.] We indeed 
see that the extrema in Fig.~\ref{fig:bulk} (b) occurring at $\ta = 0.721$
and $1.564$ agree well with some of the values of $\ta$ where the density of 
states has peaks as we have shown in Fig.~\ref{fig:dos}.
We can therefore restate the bulk-boundary correspondence by saying that the 
values of $\ta$ of the surface states match some of the points where the 
density of states has a peak in a plot like Fig.~\ref{fig:dos}.

Next, we want to see if the values of $k_z$ where the $\ta$'s of the bulk 
system have extrema have any significance for the wave functions of the surface
states. We know that the wave functions for the 1D lattice system with $N_z$ 
sites has $2N_z$ components due to the spin-1/2 degree of freedom. We therefore
consider two different Fourier transforms of the wave function $\psi_j$ of a 
surface. These are defined as
\bea f_\ua (k_z) &=& \frac{1}{N_z} \sum_{m=1,3,\cdots} e^{-ik_z m} ~\psi_j (m),
\non \\
f_\da (k_z) &=& \frac{1}{N_z} \sum_{m=2,4,\cdots} e^{-ik_z m} ~\psi_j (m), 
\non \eea
for spin-up and spin-down electrons respectively.

\begin{figure}[h]
\begin{center}
\subfigure[]{\ig[width=3.4in]{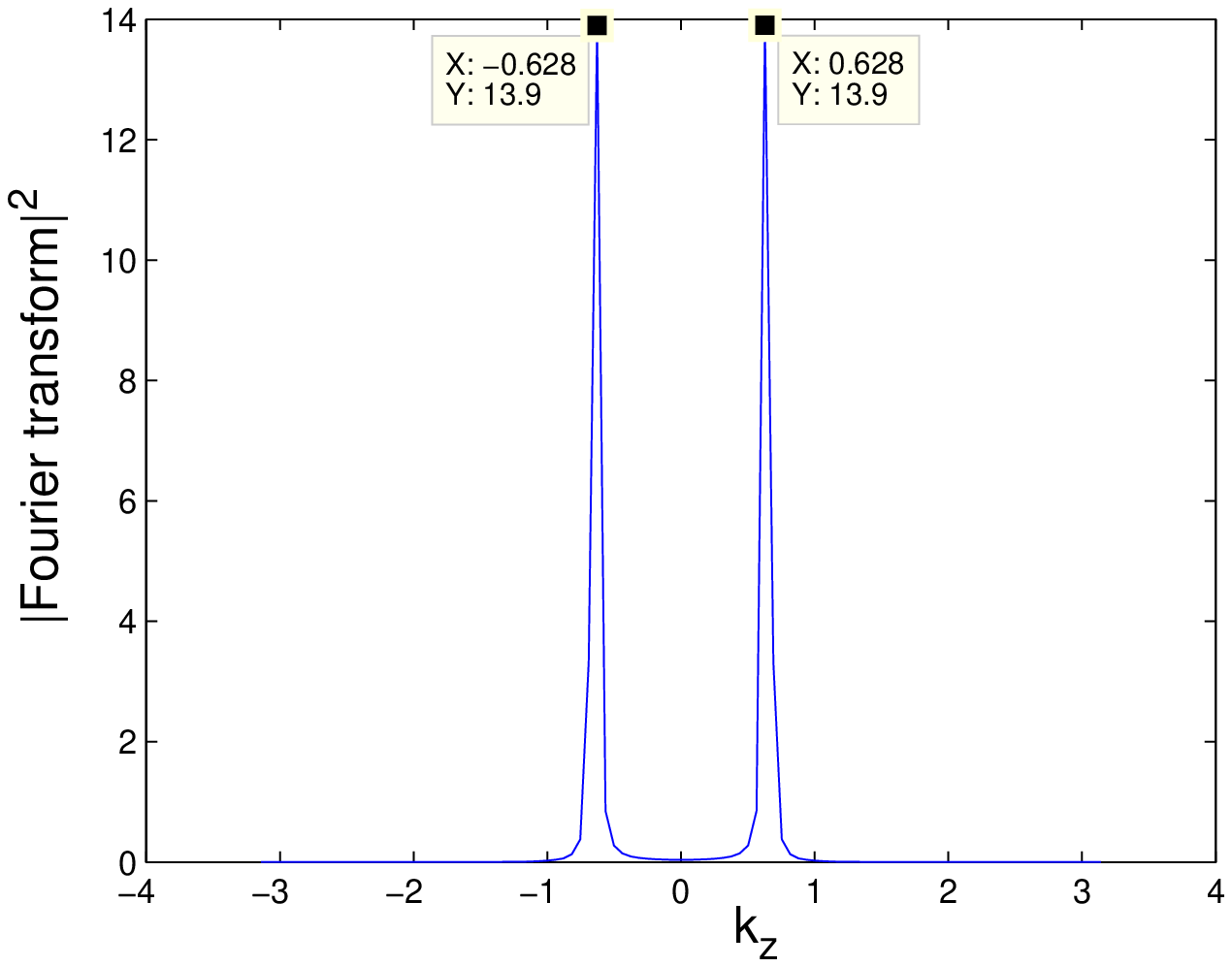}} \\
\subfigure[]{\ig[width=3.4in]{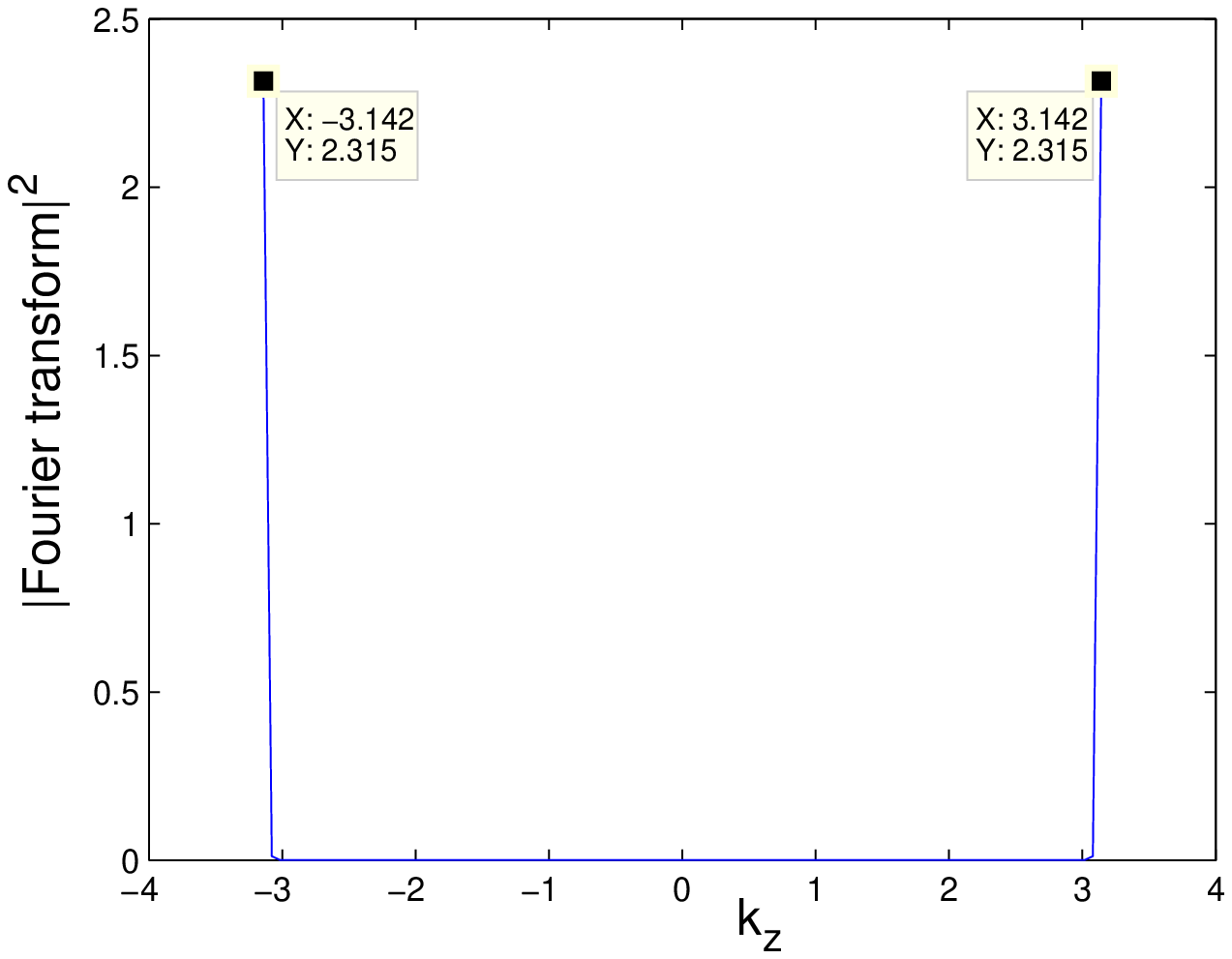}}
\end{center}
\caption{Plots of $|f_\ua (k_z)|^2$ vs $k_z$ for two surface states 
whose $\ta$'s match some of the extrema of the $\ta$'s of the bulk states. 
The coordinates of the peaks of the Fourier transforms are shown in the 
figure. The system parameters are the same as in Fig.~\ref{fig03}.} 
\label{fig:ft} \end{figure}

In Fig.~\ref{fig:ft}, we show plots of $|f_\ua (k_z)|^2$ versus $k_z$ for two 
surface states with the parameter values as shown in Fig.~\ref{fig03}; we have 
chosen surface states whose $\ta$'s match some of the extrema of the $\ta$'s of
the bulk system as shown in Fig.~\ref{fig:bulk} (b). We see that the peaks 
$|f_\ua (k_z)|^2$ occur at the values of $k_z$ equal to $\pm 0.628$ and 
$3.142$; these are the same as the values of $k_z$ where the $\ta$'s of the 
bulk system shown in Fig.~\ref{fig:bulk} (b) have extrema. (We obtain similar 
results if we plot $|f_\da (k_z)|^2$ versus $k_z$). This is another kind of 
bulk-boundary correspondence.

It would be useful to understand the reason for a correspondence between the 
extrema of the $\ta$'s of the bulk system and the $\ta$'s of the surface 
states. A possible reason may be as follows~\cite{saha2}. We know that the 
presence of an extremum of the bulk $\ta$'s near a particular value, say, 
$\ta'$ at $k_z= k'$, means that the density of states $\rho (\ta)$
diverges as we approach $\ta = \ta'$. The presence of a large
number of bulk states near $(k',\ta')$ may make it easier for a system
to superpose those states to form states which are localized at the surface.
Such surface states will then naturally have a value of $\ta$ which lies
close to $\ta'$ and a Fourier transform whose peak is close to $k'$.

Before ending this section, we would like to note that the values of $\ta$'s of
the bulk and surface states are {\it not} separated by a gap. This raises the
question, why are the surface states robust to certain kinds of perturbations, 
such as on-site disorder as discussed at the end of Sec.~\ref{sec:results1}?

\subsection{Surface states at $k_x= k_y=0$}
\label{sec:zero}

It turns out that the problem defined by Eq.~\eqref{ham3} becomes much easier 
to solve and we can gain a much better understanding of the surface state 
wave functions for the special case $k_x= k_y =0$. We then get 
\bea H\psi_n &=& (m_0 ~-~ 2m_1 ~-~ m_2 A ^2)~ \si^z ~\psi_n \non \\
&& +~ m_1 ~\si^z ~(\psi_{n+1} ~+~ \psi_{n-1}) \non \\
&& + ~v ~ (\si^x A\cos{\om t} ~+~ \si^y A\sin{\om t}) ~\psi_n, \label{Ham} \eea 
where $A = A_0 e^{- \ga a n}$.
We will write $\psi_n = (a_n, b_n)^T$, where $a_n$ and 
$b_n$ denote the spin-up and spin-down components at the site $n$. The 
Schr\"odinger equation $i \pa \psi_n /\pa t = H \psi_n$ then takes the form
\bea i\frac{\pa a_n}{\pa t} &=& (m_0 ~-~ 2 m_1 ~-~ A^2) ~a_n \non \\
&& + ~m_1 ~(a_{n+1}~+~ a_{n-1}) ~+~ v~ A e^{-i\om t} ~b_n, \non \\
i\frac{\pa b_n}{\pa t} &=& -~(m_0 ~-~ 2 m_1 ~-~ A^2) ~b_n \non\\
&& - ~m_1 ~(b_{n+1} ~+~ b_{n-1}) ~+~ v ~A e^{i\om t} ~a_n. \non \\
&& \label{ab1} \eea

Interestingly, there is a transformation which makes Eqs.~\eqref{ab1} 
look time-independent. This is given by
\bea a_n ~=~ {\tilde \al}_n ~e^{-i \om t/2} ~~~~{\rm and}~~~~ b_n ~=~ 
{\tilde \be}_n ~e^{i \om t/2}. \eea
Substituting this in Eq.~\eqref{ab1}, we obtain
\bea i\frac{\pa {\tilde \al}_n}{\pa t} &=& (m_0 ~-~ 2 m_1 ~-~ A^2 ~-~ 
\frac{\om}{2}) ~{\tilde \al}_n \non \\
&& + ~m_1 ~({\tilde \al}_{n+1} ~+~ {\tilde \al}_{n-1})~+~ v A ~{\tilde \be}_n, 
\non \\
i\frac{\pa {\tilde \be}_n }{\pa t} &=& -~(m_0 ~-~ 2 m_1 ~-~ A^2 ~-~
\frac{\om}{2}) ~{\tilde \be}_n \non \\ 
&& - ~m_1 ~({\tilde \be}_{n+1}~+~ {\tilde \be}_{n-1}) ~+~ v A ~{\tilde \al}_n. 
\label{ab2} \eea
Next, we put ${\tilde \al}_n(t)=\al_n e^{-i Et}$ and ${\tilde \be}_n(t)=\be_n 
e^{-i Et}$ in the above equations. [This implies that $a_n (T) = - a_n (0)
e^{-i E T}$ and $b_n (T) = - b_n (0) e^{-i E T}$, where we have used the fact 
that $e^{-i \om T/2} = e^{-i \pi} = -1$. Hence the FEV is given by $e^{i \ta} 
= - e^{-i E T}$.] We then find the completely time-independent equations
\bea E\al_n &=& (m_0 ~-~ 2 m_1 ~-~ A^2 ~-~ \frac{\om}{2}) ~\al_n \non \\
&& + ~m_1 (\al_{n+1} ~+~ \al_{n-1}) ~+~ v A ~\be_n \non \\
E\be_n &=& -~ (m_0 ~-~ 2 m_1 ~-~ A^2 ~-~ \frac{\om}{2}) ~\be_n \non \\
&& - ~m_1 (\be_{n+1} ~+~ \be_{n-1}) ~+~ v A ~\al_n. \label{ab3} \eea

We have numerically solved Eqs.~\eqref{ab3} for a finite system in which $n$ 
goes from 1 to $N_z$. We find several surface states $\psi_j$ which are 
localized near $n=1$ as in Figs.~\ref{fig02}-\ref{fig03}. For all these
states, we find that the Fourier transform of both the components $a_n$ and 
$b_n$ are peaked at $k_z=\pi$. This means that $\al_n e^{i \pi n} = \al_n 
(-1)^n$ varies slowly on the scale of a lattice spacing (given by $a=1$ \AA), 
and 
similarly for $\be_n (-1)^n$. We therefore define the slowly varying quantities
\bea \al'_n ~=~ (-1)^n \al_n ~~~~{\rm and}~~~~ \be'_n ~=~ (-1)^n \be_n. \eea
Eqs.~\eqref{ab3} then take the form
\bea E\al'_n &=& (m_0 ~-~ 2 m_1 ~-~ A^2 ~-~ \frac{\om}{2}) ~\al'_n \non \\
&& - ~m_1 (\al'_{n+1} ~+~ \al'_{n-1}) ~+~ v A ~\be'_n \non \\
E\be'_n &=& -~ (m_0 ~-~ 2 m_1 ~-~ A^2 ~-~ \frac{\om}{2}) ~\be'_n \non \\
&& + ~m_1 (\be'_{n+1} ~+~ \be'_{n-1}) ~+~ v A ~\al'_n. \label{ab4} \eea


Since $\al'_n, ~\be'_n$ are slowly varying, we can use the Taylor expansions
$\al'_{n+1}+\al'_{n-1} = 2 \al'_n + \pa^2 \al'_n /\pa z^2$ and 
$\be'_{n+1}+\be'_{n-1} = 2 \be'_n + \pa^2 \be'_n /\pa z^2$, where $z = -na$ is
now a continuous variable. [We have taken $z=-na$ because $n$ increases from 
zero while $z$ decreases from zero as we go down from the top surface;
see Fig.~\ref{fig:weyl}. Hence $A=A_0 e^{-\ga a n} = A_0 e^{\ga z}$.]
We will henceforth write $\al'_n = \al' (z)$ and 
$\be'_n = \be' (z)$. Using this in Eqs.~\eqref{ab4} gives
\bea E\al' &=& (m_0 ~-~ 4 m_1 ~-~ A^2 ~-~ \frac{\om}{2}) ~\al' \non \\
&& - ~m_1 ~\frac{\pa^2 \al'}{\pa z^2} ~+~ v A ~\be', \label{ab5} \eea
\bea
E\be' &=& - ~(m_0 ~-~ 4 m_1 ~-~ A^2 ~-~ \frac{\om}{2}) ~\be' \non \\
&& + ~m_1 ~\frac{\pa^2 \be'}{\pa z^2} ~+~ v A ~\al'. \label{ab6} \eea

Next, we find numerically that for the surface states, $E \simeq m_0-4m_1 -
\om/2$, so that 
\beq \De E ~=~ E - (m_0-4m_1- \om/2) ~\ll~ E. \label{de} \eeq
We also find that $\be'$ is much smaller than $\al'$ for the surface states; 
hence we can neglect the second order differential term in Eq.~\eqref{ab6}. 
We can also ignore the term $A^2 = A_0^2 e^{2 \ga z}$ in Eq.~\eqref{ab6} if 
$A_0$ is small. That equation then gives $E\be' \simeq - E\be' + v A\al'$ 
which implies
\beq \be' ~=~ \frac{v A \al'}{2E}. \label{ben} \eeq
Using Eqs.~(\ref{de}-\ref{ben}) in Eq.~\eqref{ab5}, we get
\beq - ~m_1 ~\frac{\pa^2 \al'}{\pa z^2} ~+~ \frac{v^2 A^2}{2E} ~\al' 
~-~ A^2 \al' ~=~ \De E ~\al'. \eeq

We now use the form $A = A_0 e^{\ga z}$. Further, since $\ga$ is small and we
are only interested in the region close to $z=0$ (the top surface), we can
write $A = A_0 ( 1 + \ga z)$. Putting all this together, we obtain
\beq - \frac{\pa^2 \al'}{\pa z^2} ~-~ \frac{A_0^2}{m_1} ~(1 - \frac{v^2}
{2E})~ (1 + 2 \ga z) ~\al' ~=~ \frac{\De E}{m_1} ~\al'. \label{al1} \eeq
For the parameter values that we are using, $E \simeq m_0-4m_1 - \om/2$ is
negative. Hence the quantity
\beq C ~=~ \frac{2}{m_1} ~(1 - \frac{v^2} {2E}) \eeq
is positive. Defining
\beq D ~=~ \frac{\De E}{m_1} ~+~ \frac{A_0^2}{m_1} ~(1 - \frac{v^2} {2E}), \eeq
Eq.~\eqref{al1} takes the form
\beq - \frac{\pa^2 \al'}{\pa z^2} ~-~ C \ga A_0^2 z \al' ~=~ D \al'. 
\label{airy1} \eeq
This is the Schr\"odinger equation of a particle in a potential
which increases linearly as $z$ decreases from zero, and there is a 
hard wall at $z=0$ since the particle cannot be in the region $z > 0$
(which lies outside system).

We now define a rescaled variable
\beq w ~=~ (C \ga A_0^2)^{1/3} ~\left( z ~+~ \frac{D}{C \ga A_0^2} \right). 
\label{zw} \eeq
Eq.~\eqref{airy1} then takes the form
\beq - \frac{\pa^2 \al'}{\pa w^2} ~-~ w \al'_n ~=~ 0. \label{airy2} \eeq
This is known as the Airy equation; its solutions are given
by the first and second kind of Airy functions denoted by $Ai(-w)$ and 
$Bi(-w)$ respectively. We want a solution of Eq.~\eqref{airy2} which
goes to zero as $w \to - \infty$; such a solution is given by $Ai(-w)$ which
has the asymptotic form $Ai(-w) \sim \exp [-(2/3) (- w)^{3/2}]$ as $w \to - 
\infty$. In general a linearly increasing potential gives several bound 
states. The spreads of the wave functions of these bound states will be 
given by $\De w$ of order 1, which implies
\beq \De z ~\sim~ \frac{1}{\ga^{1/3} ~A_0^{2/3}}, \label{dez} \eeq
due to the form in Eq.~\eqref{zw}.

We thus conclude that for a fixed amplitude of the radiation, $A_0$, the
spread of the bound states is of order $a^{2/3}/\ga^{1/3}$ (we have introduced
the factor of $a^{2/3}$ for dimensional reasons), while the 
penetration length of the radiation is $1/\ga$. Clearly, $1/\ga$ is much 
larger than $a^{2/3}/ \ga^{1/3}$ if $1/(a\ga) \gg 1$. It is therefore possible 
to choose $\ga$ and the system size $N_z a$ in such a way that the spreads of 
the bound states are much smaller than $N_z a$ which, in turn, is much smaller 
than $1/\ga$. This is the regime in which all our numerical calculations have 
been done. Note that since we have taken $N_z a \ll 1/\ga$ for the system with 
surfaces in Sec.~\ref{sec:results1}, we are 
justified in doing the calculations for the bulk system in Sec.~\ref{sec:bbc} 
with $\ga$ set equal to 0; this is necessary since we want the bulk system 
(with periodic boundary conditions) to be translation invariant in the
$\hat z$-direction.

Eq.~\eqref{dez} implies that if $\ga$ is held fixed and the radiation amplitude
$A_0 \to 0$, the spreads of the bound states diverge. This makes sense
since we do not expect to find any bound states when $A_0 = 0$.

\begin{figure}[h!]
\begin{center}
\subfigure[]{\includegraphics[width=4.2cm]{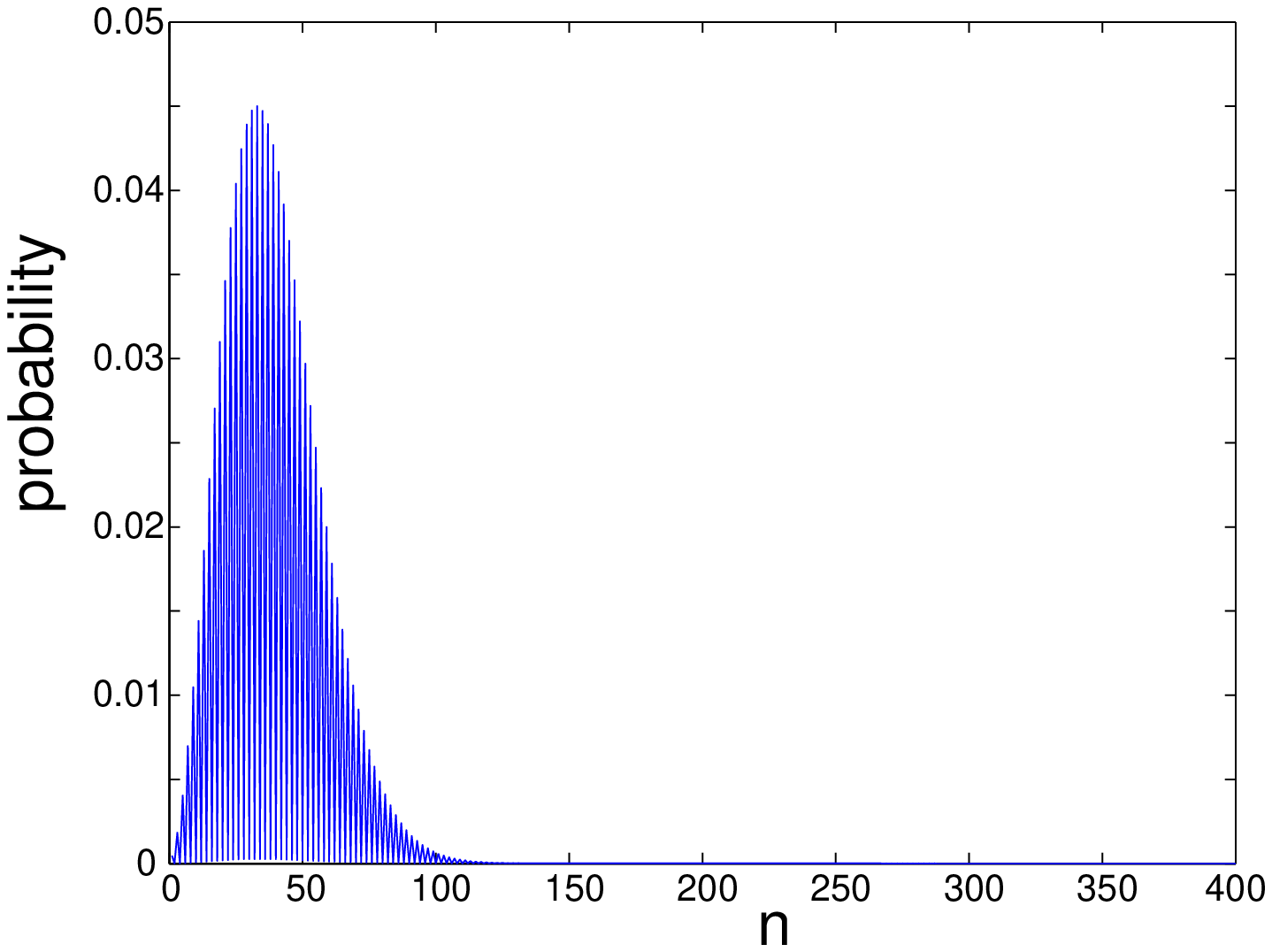}}
\subfigure[]{\includegraphics[width=4.2cm]{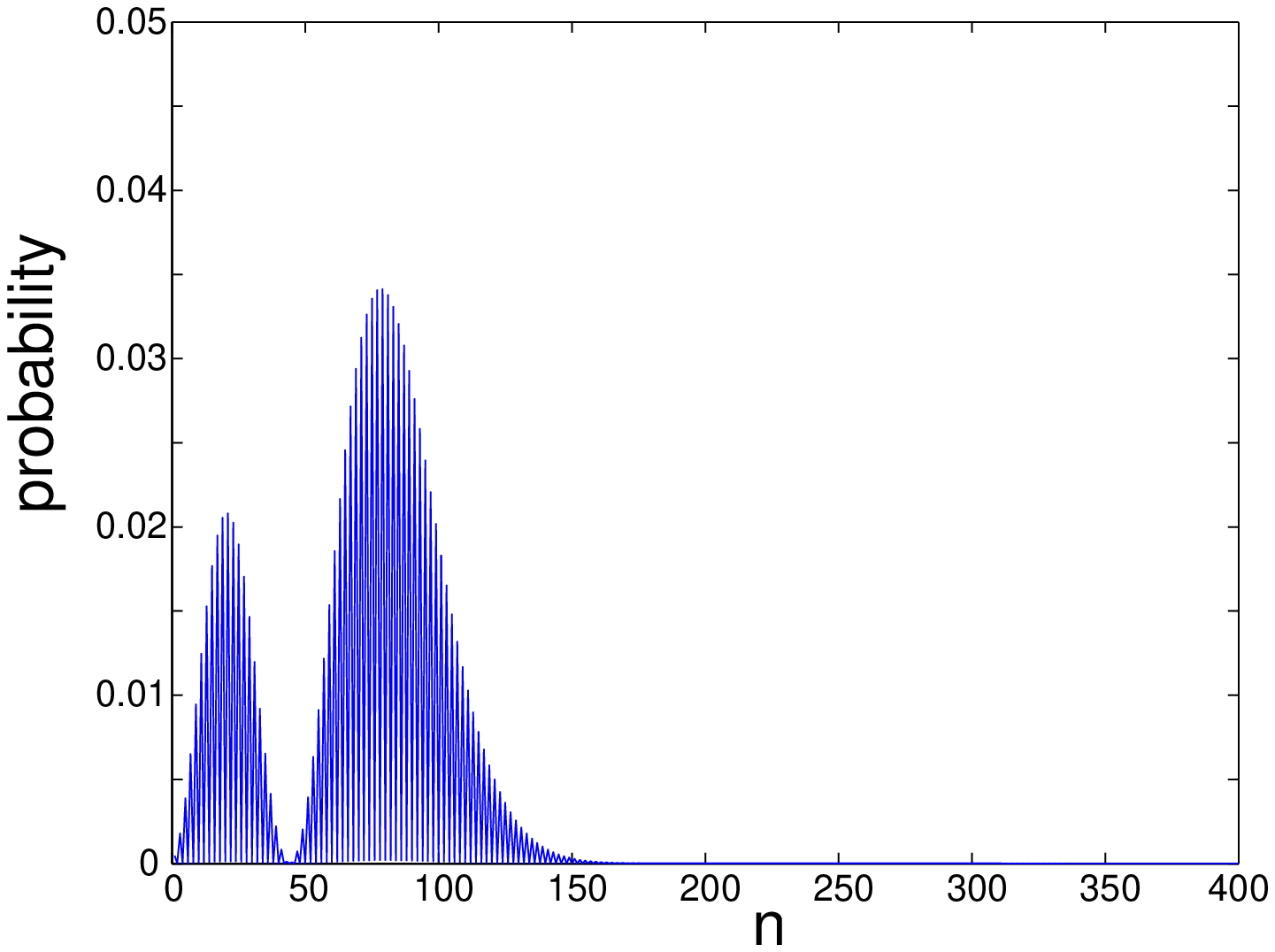}} \\
\subfigure[]{\includegraphics[width=4.2cm]{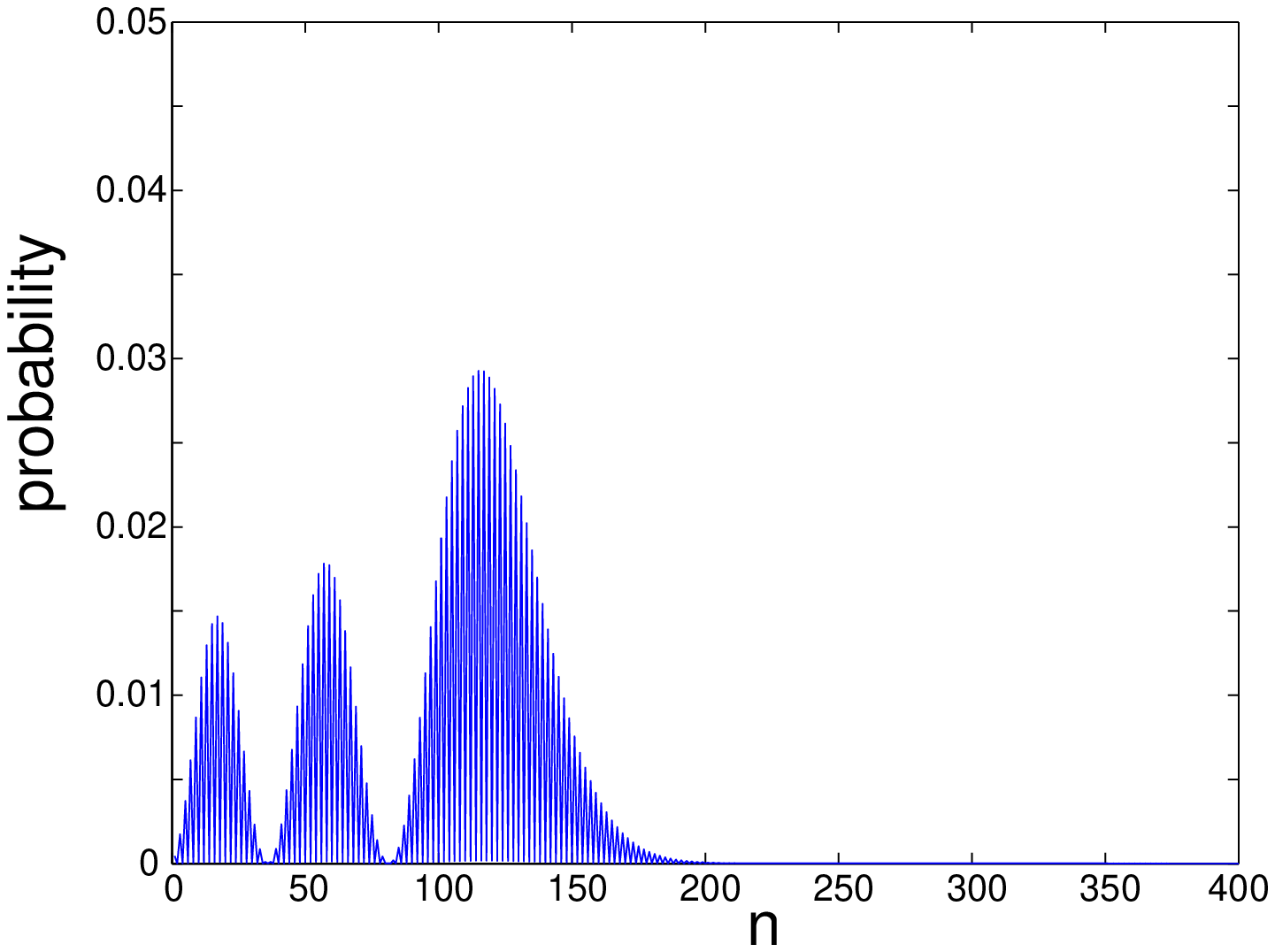}}
\subfigure[]{\includegraphics[width=4.2cm]{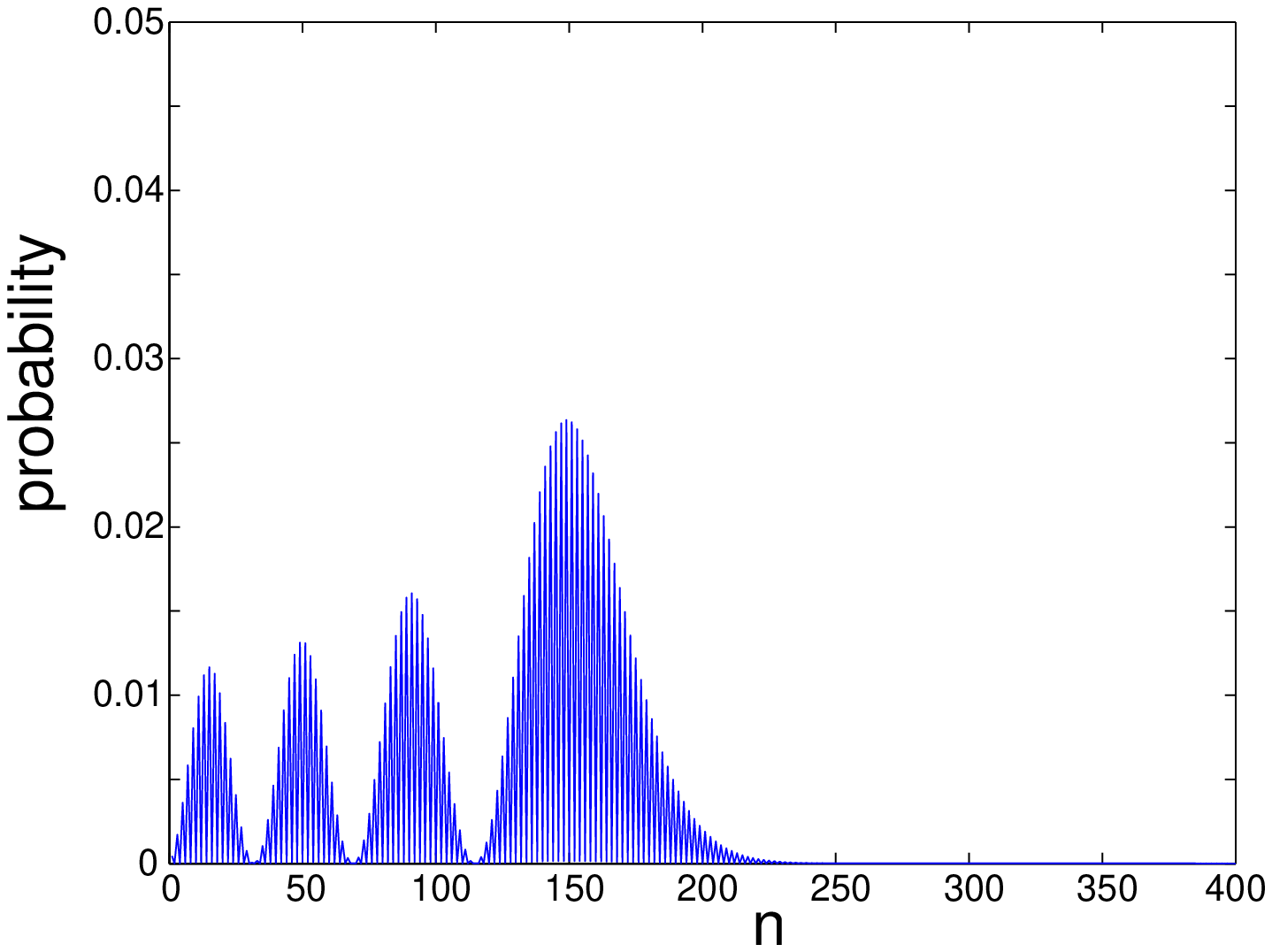}} \\
\caption{Probabilities vs $n$ for four surface state wave functions for 
a system with $m_0=0.5$ eV, $m_1= 0.605$ eV, $m_2=1$ eV \AA$^2$, $v=1$ eV \AA,
$k_x = k_y = 0$, $A_0=0.5$ \AA$^{-1}$, $\ga=0.0005$ \AA$^{-1}$, $\om=2$ 
eV$/\hbar$, and 200 sites.} \label{fig:airy} \end{center} \end{figure}

Fig.~\ref{fig:airy} shows four surface states with 0, 1, 2 and 3 nodes
respectively, for a system with $m_0=0.5$ eV, $m_1=0.605$ eV, $m_2=1$ eV 
\AA$^2$, $v=1$ eV \AA, $k_x=k_y = 0$, $A_0=0.5$ \AA$^{-1}$, $\ga=0.0005$ 
\AA$^{-1}$, $\om=2$ eV$/\hbar$, and 200 sites.

Finally, we note that although we have derived the scaling relation in 
Eq.~\eqref{dez} for the case $k_x = k_ y = 0$, we expect this relation to 
hold even for other momenta as long as the amplitude $A_0$ of the EMR is small.
A comparison of Figs.~\ref{fig02} and \ref{fig03}, where $k = 0.5$ \AA$^{-1}$, 
shows that the spread of the bound state indeed increases when $A_0$ decreases.



\section{Effects of periodic driving on topological insulators}
\label{sec:topo}

\subsection{Hamiltonian}
\label{sec:model2}

We now study what happens when EMR applied to the top surface of a TI like 
$Bi_2 Se_3$~\cite{wang3}. We begin with a bulk Hamiltonian for $Bi_2 Se_3$ 
of the form~\cite{qi} 
\bea H &=& (-m+B_1 k_z^2) \tau^z ~+~ v_z k_z \tau^y \non \\
&&+~ v_\parallel \tau^x (\si^x k_y - \si^y k_x), \label{ham6} \eea
where $m=0.28$ eV, $B_1=6.86$ eV \AA$^2$, $v_z=2.26$ eV \AA, and 
$v_\parallel=3.33$ eV \AA. Here the $\si^i$'s are Pauli spin matrices, and
$\tau^i$'s denote Pauli pseudospin matrices with $\tau^z=\pm 1$ denoting 
$Bi$ and $Se$ respectively. In the presence of circularly polarized EMR 
applied to the top surface ($x-y$ surface located at $z=0$), the 
Hamiltonian becomes,
\bea H &=& (-m+B_1 k_z^2)\tau^z + v_z k_z \tau^y \non \\
&& + v_\parallel \tau^x [\si^x (k_y+A \sin(\om t)) -\si^y (k_x+A\cos(\om t))],
\non \\
&& \label{ham7} \eea
where we assume the form $A = A_0 e^{\ga z}$. As discussed in 
Sec.~\ref{sec:model1}, we again introduce a 1D lattice, with a lattice spacing 
$a =1$ \AA, along the $\hat z$-direction. Each lattice point now has four 
components corresponding to spin $\si^z=\pm 1$ and pseudospin $\tau^z=\pm 
1$~\cite{deb}. To go from the continuum Hamiltonian in Eq.~\eqref{ham7} to a 
lattice Hamiltonian, we replace $k_z^2 \to (2/a^2 )[1 - \cos(k_z a)]=(2/a^2)
(1-(e^{ik_za}+e^{-ik_za}) /2)$, and $k_z \to (1/a)[\sin (k_z a)]=(-i/a) 
[e^{ik_za}+e^{-ik_za}]$, assuming that $k_z a \ll 1$. We then set $a=1$ \AA~
as before. Following these substitutions, we obtain the action of the lattice
Hamiltonian on the four-component wave function $\psi_n$
\bea H\psi_n &=& -~m \tau^z \psi_n ~+~ 2B_1 \tau^z \psi_n ~-~ B_1 \tau^z
(\psi_{n+1}+\psi_{n-1}) \non \\
&&- ~\frac{i}{2} ~v_z \tau^y (\psi_{n+1}-\psi_{n-1}) \non \\
&&+ ~v_\parallel \tau^x [\si^x (k_y ~+~ A\sin(\om t)) \non \\
&& ~~~~~~~~~~- ~\si^y (k_x ~+~ A\cos(\om t))] \psi_n, \label{ham8} \eea
where $A = A_0 e^{-\ga n}$. As we discussed for the WSM, this Hamiltonian is 
again rotationally invariant; hence we can set $k_y=0$ henceforth.

\subsection{Numerical Results}
\label{sec:results2}

We now numerically compute the Floquet operator
\beq U ~=~ {\cal T} e^{-i \int_0^T dt H(t)}, \eeq
where $T = 2\pi/\om$, and find its eigenvalues and eigenstates. We choose 
$k_x = k =0.5$ \AA$^{-1}$ and $k_y = 0$, and take the parameters of the EMR 
to be $A_0 =1$ \AA$^{-1}$, $\ga=0.001$ \AA$^{-1}$, $\om=2$ 
eV$/\hbar$, and $N_z=200$ sites ($H$ and $U$ are therefore 800-dimensional
matrices). We know that the $x-y$ surface of a TI hosts surface states even 
in the absence of any periodic driving. We therefore want to see whether the
EMR generates any new surface states.

\vspace{.5cm}
\begin{figure}[h]
\epsfig{figure=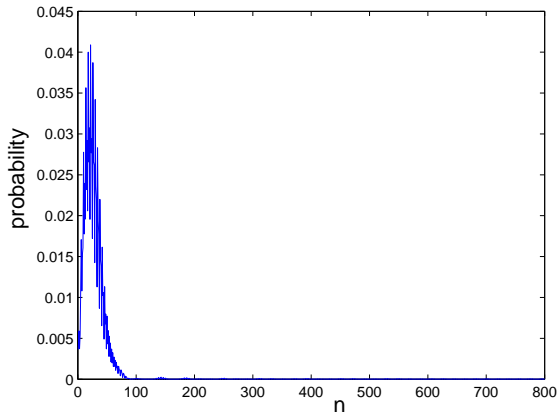,width=8cm}
\caption{Probabilities vs $n$ for a state on the $x-y$ surface of a TI 
with a small decay length. The system parameters are $m=0.28$ eV, $B_1=6.86$ 
eV \AA$^2$, $v_z=2.26$ eV \AA, $v_\parallel =3.33$ eV \AA, $A_0 =1$ \AA$^{-1}$, 
$\ga=0.001$ \AA$^{-1}$, $\om=2$ eV$/\hbar$, and 200 sites.} \label{fig:sf1} 
\end{figure}

\begin{figure}[h]
\epsfig{figure=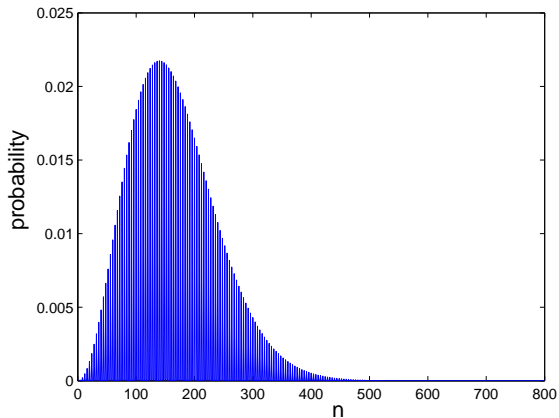,width=8cm}
\caption{Probabilities vs $n$ for a state on the $x-y$ surface of a TI 
with a decay length which is much larger than the one in Fig.~\ref{fig:sf1}.
The system parameters are the same as in Fig.~\ref{fig:sf1}.} \label{fig:sf2} 
\end{figure}

We find that there are two types of surface states on the $x-y$
surface. There are some surface states which appear even in the absence
of driving and have a decay length given by $v_z/m \sim 8$ \AA; we call
these the old surface states. Figure~\ref{fig:sf1} shows this kind of a 
surface state. The decay lengths of these states are almost independent of
the radiation amplitude. Figure~\ref{fig:sf2} shows another kind of surface 
state which has a much larger decay length. These are the new states that 
appear due to periodic driving by the EMR. We find that the decay lengths of 
these states vary with the amplitude of the EMR just as we found for the 
surface states of a WSM in Sec.~\ref{sec:results1}.


If the value of the parameter $m$ in Eqs.~(\ref{ham6}-\ref{ham7}) is changed
from $0.28$ eV to $- 0.28$ eV, the system becomes a trivial insulator.
When a circularly polarized EMR is applied to the $x-y$ surface of this
system, we find numerically that the new kind of surface state with 
a large decay length again appears. However, the old states with a small \
decay length does not appear; this is expected since a trivial insulator
does not have any surface states in the absence of driving.

We have studied if a bulk-boundary correspondence exists for this system 
(either a TI or a trivial insulator) which has the same form as the
correspondence that we found in Sec.~\ref{sec:bbc}. We did not find such
a correspondence here. This may be because this system is more complicated
than a WSM in that the wave function has four components instead of two
components at each site.


%


Just as in Sec.~\ref{sec:zero}, we find that it is much easier to study 
surface states in the special case where $k_x = k_y = 0$. Once again, there 
is a unitary transformation which maps the Floquet problem to one in which 
we have to solve a time-independent problem. Namely, if we define
\beq \psi_n ~=~ \exp [-\frac{i}{2} \om t \si^z ] {\tilde \psi}_n, \eeq
followed by ${\tilde \psi}_n (t) = {\tilde \psi}_n e^{-iEt}$, we obtain 
the equation
\bea && - m \tau^z {\tilde \psi}_n ~+~ 2B_1 \tau^z {\tilde \psi}_n - B_1 \tau^z
({\tilde \psi}_{n+1} + {\tilde \psi}_{n-1}) \non \\
&& -~ \frac{i}{2} ~v_z \tau^y ({\tilde \psi}_{n+1} - {\tilde \psi}_{n-1}) 
~-~ v_\parallel \tau^x \si^y A_0 e^{-\ga n} {\tilde \psi}_n \non \\
&& -~ \frac{\om}{2} \si^z {\tilde \psi}_n ~=~ E {\tilde \psi}_n. 
\label{psiE} \eea
This reproduces the solutions of the Floquet problem, with the Floquet
eigenvalues being given by $e^{i \ta} = - e^{-iET}$. As before the surface 
states can be found by looking at the IPRs of all the solutions of 
Eq.~\eqref{psiE} and identifying the ones with the largest IPRs. 

\section{Conclusions}
\label{sec:con}

In this paper, we have studied the effects of EMR applied to the top surface 
of a WSM. We first introduced a simple model for a WSM which has only two
Weyl points. The projections of these Weyl points on the different surfaces 
give the end points of Fermi arcs; these are curves lying in the Brillouin 
zones of those surfaces such that there are surface states whose momenta lie 
on the Fermi arcs. For the top surface of the WSM in our model, the Fermi arc 
is a single point; hence there is Fermi arc and no state on that surface in 
the time-independent system where the EMR is absent. We then studied the 
system in the presence of circularly polarized EMR applied to the top surface.
We choose the case of circular polarization so that the Hamiltonian 
is invariant under rotations within the top surface. 

To numerically look for surface states, we introduce a 1D lattice along the 
$\hat z$-direction. At each lattice point, denoted by an integer $n$, 
there are two states corresponding to the electron spin, $\si^z=\pm 1$. 
We take the amplitude of the vector potential of the EMR
to be of the form $A=A_0 e^{-\ga a n}$, where $1/\ga$ is the penetration 
length of the radiation. It is realistic to assume this kind of exponential 
decay as radiation usually gets absorbed in a medium; we also find that this 
assumption is necessary for our calculations as surface states do not appear 
if we take an infinitely large penetration length. Due to the translational 
invariance of the surface, the momentum on the surface,
$(k_x,k_y)$, is a good quantum number. We numerically calculate the Floquet 
operator $U$ which evolves the system through one time period $T=2 \pi/\om$,
where $\om$ is the frequency of the EMR. Using the rotational symmetry, we 
have shown that the FEVs do not depend on $k_x$ and $k_y$ separately, but 
only on the magnitude $k=\sqrt{k_x^2+k_y^2}$. Hence we have assumed
$k_x=k$ and $k_y=0$ in all our calculations.

Given all the eigenstates of $U$, we find which of them describe 
surface states by calculating the IPRs of all the states and identifying the
ones with large values of IPR. The surface states decay rapidly as we go into 
the bulk. Looking at the surface states for different values of the amplitude
$A_0$ (specifically, $A_0 = 0.05$ and $0.5$), we find that the 
decay length of the surface state is larger if $A_0$ is smaller.
We show later that there is a power law which relates the two quantities.
We find that the number of surface states is proportional to the area of
the surface.

For time-independent systems, it is known that a non-trivial bulk topology 
gives rise to states at the surfaces of the system. Hence there is a 
correspondence between the bulk bands and the surface states. We look for 
such a bulk-boundary correspondence in our system. To do this, we consider a 
bulk system with periodic boundary conditions along the $\hat z$-direction so 
that all the three momenta, $k_x$, $k_y$ and $k_z$, are good quantum numbers. 
We then plot $\ta$ (corresponding to the FEVs $e^{i\ta}$) of the bulk system 
as a function of $k_z$, for a given value of $k_x = k$ and $k_y =0$. We find 
that the values of $\ta$ of the surface states match some of the extrema of 
the $\ta$'s of the bulk states. To put this differently, the density of 
states $\rho (\ta)$ of the bulk system has peaks at certain values of $\ta$,
and we find that some of these points match the $\ta$'s of the surface states.
Next, we find the Fourier transform $f(k_z)$ of the surface states as a 
function of $k_z$, for one particular spin component. We find that the peaks 
of $|f(k_z)|^2$ occur at the same values of $k_z$ where the $\ta$'s of the 
bulk system have extrema. We therefore find two kinds of bulk-boundary 
correspondence, one between the values of $\ta$'s of the bulk and surface 
states and the other between the corresponding values of $k_z$.

We then study a special case of this problem where $k_x= k_y = 0$. We find 
that this problem can be made completely time-independent by a unitary 
transformation. It is possible to solve the problem analytically in 
the continuum limit; we obtain a differential equation of the Airy form
which explains why the numerically obtained surface state wave functions
look like Airy functions. We find that the spread of these bound states scales
with the amplitude $A_0$ and the penetration length $1/\ga$ of the EMR
as $1/(\ga^{1/3} A_0^{2/3})$. Thus the spread of the bound states diverges
if either $\ga$ or $A_0$ goes to zero.

Finally, we study the effect of circularly polarized EMR on a TI. A TI
has non-trivial topology in the bulk and hence hosts states at the surfaces 
even in the absence of EMR. We looked at the effects of the EMR to see if this
generates a new kind of surface states. A numerical study of the 
Floquet operator $U$ shows that there are two kinds of surface states; one 
is the usual (old) surface state found in the time-independent system while 
the other is a new surface state whose decay length is much larger than that
of the old surface state. The decay lengths of the new surface states
can be tuned by varying the parameters of the EMR. We then find that if we 
consider a trivial insulator, there is no surface state in the absence 
of the EMR as expected; however, the application of EMR generates surface 
states which are similar to the new surface states generated in a TI.
In analogy with the WSM, we have looked for a bulk-boundary correspondence
for periodic driving of a TI, but have not found such a correspondence so far.

Turning to experimental methods for detecting the surface states
generated by EMR, it may be possible to look for such states using
time-resolved and angle-resolved photoemission spectroscopy~\cite{wang3}
and anomalous Hall conductivity~\cite{chan}.
One can also try to measure the oscillating surface currents which are 
carried by the surface states~\cite{gonzalez}.

We will end by pointing out some directions for future studies. In this 
paper, we have only considered circularly polarized EMR; this simplifies the 
calculations due to rotational invariance. We can study what happens if the 
EMR is linearly polarized; in this case, the form of the surface states will 
depend on both $k_x$ and $k_y$, not just $\sqrt{k_x^2 + k_y^2}$. It would be 
interesting to find the complete range of parameters (such as the frequency 
$\om$ of the EMR and the surface momentum $(k_x,k_y)$) in which surface states
appear. The stability of the surface states may be worth studying in detail.
References~\onlinecite{gonzalez,gonzalez2} have argued that there is a 
topological 
protection because the momenta of the surface states are given by exceptional 
points which have a branch point structure. However, it would be useful to 
see if any topological invariants appear in this problem; this may help to 
analytically find the values of $\om$ where surface states appear or 
disappear~\cite{saha2,kitagawa2,thakurathi}. For the WSM, we have 
only studied the states on the top surface ($x-y$) because we know that 
there are no states there in the absence of EMR. It would be interesting
to study what happens on the other surfaces where states are present even
when there is no EMR~\cite{gonzalez}.

\vspace*{.5cm}
\section*{Acknowledgments}

D.S. thanks Department of Science and Technology, India for 
Project No. SR/S2/JCB-44/2010 for financial support.

\end{document}